\documentclass[aps,prl,reprint,groupedaddress]{revtex4-1}
\usepackage[dvipdfm]{graphicx}

\begin{document}


\title{Effects of error on fluctuations under feedback control}


\author{Sosuke Ito and Masaki Sano}
\affiliation{Department of Physics, The University of Tokyo - Hongo, Bunkyo-ku, Tokyo, Japan}


\date{\today}

\begin{abstract}
We consider a one-dimensional Brownian motion under nonequilibrium feedback control. Generally, the fluctuation-dissipation theorem (FDT) is violated in driven systems under nonequilibrium conditions. We find that the degree of the FDT violation is bounded by the mutual information obtained by the feedback system when the feedback protocol includes measurement errors.  We introduce two simple models to illustrate cooling processes by feedback control and demonstrate analytical results for the cooling limit in those systems.  Especially in a steady state, lower bounds to the effective temperature are given by an inequality similar to the Carnot efficiency.
\end{abstract}


\maketitle

\section{Introduction}
Discussions on the Maxwell's demon have provided a better understanding of the relation between information entropy and entropy production~\cite{Szilard,demon,Maruyama}. As a generalization of the relation between information entropy and entropy production, the second 
law of thermodynamics is extended to an open system under feedback control~\cite{Sagawa,Sagawa2}. The generalization of the second law is denoted by
\begin{equation}
\beta \left( \left< W \right> - \Delta F \right) \geq - \left<I \right>,
\label{SagawaUeda}
\end{equation}
where $\left< W \right>$ is the ensemble average of work $W$ exerted on the open system, $\Delta F$ is the free energy difference gained in the open system, and $\left< I \right>$ is the mutual information obtained by the feedback protocol. The open system is in contact with a thermal reservoir at temperature $T =(k_{\rm B} \beta)^{-1}$, where $k_{\rm B}$ is the Boltzmann constant. The difference $\left< W \right> - \Delta F $ amounts to a dissipated work in the open system. When the dissipated work becomes negative, the feedback can extract work from a heat reservoir. The amount of work is bounded by the mutual information $\left< I \right>$, owing to the generalized second law, Eq.(\ref{SagawaUeda}). 

Feedback control in Brownian systems has important applications in noise cancellation, namely, cold damping or entropy pumping~\cite{Kim,Jourdan}. For instance in cold damping, thermal noise of the cantilever in an atomic force microscope was canceled through a measurement of velocity and feedback control with a force proportional to the velocity of the cantilever~\cite{Jourdan}. Similarly, in entropy pumping, reduction of thermal fluctuations by optical tweezers under velocity-dependent feedback control was proposed~\cite{Kim}.
These discussions did not take into account noise effects in the feedback system, which are unavoidable in a real experiment. An ideal condition that the effective temperature reaches 0 K was only discussed in Ref.~\cite{Jourdan}. The fundamental limit of cooling by feedback in the presence of measurement errors has not been discussed. 

 To discuss the noise effects, we study the generalized second law for a one-dimensional Langevin system and derive the relation between fluctuations and mutual information. In our derivation, we can apply the following remarkable progress in nonequilibrum statistical mechanics. The fluctuation theorem (FT)~\cite{Evans,Gallavotti,Crooks} and the Jarzynski equality~\cite{Jarzynski} are remarkable advances which are connected to the second law. The premise of the FT, the detailed FT, which is the FT for specific trajectory, is also the premise of the generalized second law~\cite{Sagawa2}. The detailed FT can be derived for many systems including a Langevin system~\cite{Evans2,Jarzynski2,Chernyak}. Maxwell's demon can be discussed using the FT for a Langevin system~\cite{Kim}. Moreover there are relations between fluctuations and entropy change for a Langevin system. The Harada-Sasa equality or the generalization of the fluctuation-dissipation theorem (FDT)~\cite{Harada,Harada2,Speck} clarifies the relations between the rate of energy dissipation and the violation of the FDT. The FDT is the relations between thermal fluctuations and the dissipation in equilibrium and is connected to the FT and the Jarzynski equality~\cite{Evans2, Jarzynski}. Generalizations of the FDT for nonequilibrium processes can generally be obtained by the perturbation dependence of a path probability~\cite{Harada2, Baiesi}.

In our discussion, we derive the Harada-Sasa equality and the generalized second law for a nonequilibrium transition performed by feedback control. Since these two equalities are connected in terms of the entropy change in the heat reservoir, we can obtain the bounds to the FDT violation. The FDT violation is bounded by the mutual information characterized as measurement errors of the feedback system. Hence, the expression of the bounds quantifies the effects of error on the FDT violation.  Here we show that effects of error are dominant especially in a cold damping system. We construct two cold damping models under velocity-dependent feedback control including measurement errors and discuss the effects of error on the FDT violation. 
Furthermore, in view of the effective temperature, the bounds to the FDT violation give the cooling limit of the effective temperature in a steady state. The lower bound to the effective temperature is determined by the balance between the information obtained by the measurement for feedback control and the information lost as a result of the relaxation. The inequality giving the lower bound to the effective temperature has a form similar to that of the Carnot efficiency.
\section{system and feedback protocol} 
We study an underdamped Langevin equation including the feedback described as
\begin{equation}
m\ddot{x}(t) + \gamma \dot{x}(t) = F_{\lambda(t,y)}(x(t)) + \epsilon f_p(t) + \xi(t),\\
\label{Feedback langevin eq}
\end{equation}
where $m$ is the mass of a Brownian particle and $\gamma$ is the friction coefficient. We assume that the friction coefficient $\gamma$ does not depend on the time $t$. The feedback force $F_{\lambda(t,y)}(x(t))$ is the external force, which generally includes a potential force $ -\partial U/\partial x$, and a constant driving force $f_{\rm ex}$ as in Ref.~\cite{Harada}. $\lambda(t,y)$ is a control parameter for a nonequilibrium transition which depends on the time $t$ and measurement outcomes $y=\{ y_1, \dots ,y_n \}$. $\epsilon f_p (t)$ is the perturbation force which is introduced for the discussion of the response function. The thermal noise $\xi(t)$ is zero-mean white Gaussian noise with variance $2\gamma k_{\rm B} T$. Throughout our paper, multiplications of stochastic variables are assumed to be interpreted as the Stratonovich-type integral without explicit remarks.

 We consider a nonequilibrium transition performed by the feedback force $F_{\lambda(t,y)}(x(t))$ from time $t=0$ to $t= \tau$. We note the phase space point of the Langevin system at time $t$ as $\Gamma(t)=(x(t),\dot{x}(t))$ and the trajectory of a transition as $\hat{\Gamma} = \{\Gamma(t)|0 \leq t \leq \tau \}$. We assume that measurements for the feedback control are performed at time $t = t_{M_i}$ $(i=1,\dots,n)$, where $0\leq t_{M_1}\leq \dots \leq t_{M_{n}} \leq\tau$, and the measurement outcomes $y_i$ are obtained at time $t = t_{M_i}$. The probability of obtaining the measurement outcome $y_i$ is depend on the phase space point $\Gamma_{M_i} = \Gamma(t_{M_i})$. Therefore the stochastic process of measurement outcomes $y$ is determined by conditional probabilities
\begin{equation}
p_i(y_i| \Gamma_{M_i} ) = g_i(y_i,  \Gamma_{M_i} ),
\label{conditional probability eq}
\end{equation}
where $g_i(y_i,  \Gamma_{M_i} ) $ is a function which characterizes the measurement error of the feedback system.
The conditional probabilities are normalized as $\int dy_i p_i(y_i|\Gamma_{M_i})=1$. 

\begin{figure}
\includegraphics[scale=0.6]{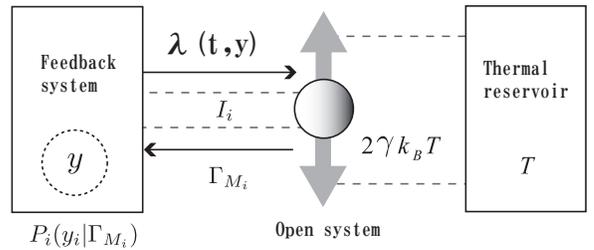}%
\caption{\label{fig.0} An open system subjected to thermal noise is coupled to a feedback system. The measurement outcome $y_i$ is obtained as a function of the phase space $\Gamma_{M_{i}}$. The control parameter $\lambda$ depends on the measurement outcomes $y$ under feedback control. $I_i$ is the mutual information which is characterized as the dependence between the measurement outcome $y_i$ and the phase space $\Gamma_{M_{i}}$. $2\gamma k_{\rm B} T$ is the value of the thermal fluctuation.}
\end{figure}

In the system, the measurement outcomes $y$ determine the time evolution of the feedback force $F_{\lambda(t,y)}(x(t))$. Due to the causality of feedback control, the time evolution of the feedback force $F_{\lambda(t,y)}(x(t))$ depends on the $i$-th measurement outcome $y_i$ for $t\geq t_{M_i}$ (see Fig.~\ref{fig.0}). 

When the  measurement outcomes $y$ are fixed, the time evolution of the control parameter $\lambda(t,y)$ is uniquely determined. Then the path probability $\mathcal{P}^\epsilon_{\lambda(t,y)}[\hat{\Gamma}| \Gamma(0)]$ for the Langevin equation Eq.(\ref{Feedback langevin eq}), is given by the Stratonovich-type path-integral expression as
\begin{equation}
\mathcal{P}^\epsilon_{\lambda(t,y)}[\hat{\Gamma}| \Gamma(0)] = \frac{1}{\mathcal{N}} e^{-\frac{\beta}{4\gamma} \int^{\tau}_0 dt  \left( m \ddot{x} + \gamma \dot{x} -F_{\lambda} -\epsilon f_p \right)^2 },
\label{path integral expression}
\end{equation}
where $\mathcal{N}$ is a normalization constant independent of $\epsilon$. The path probability $\mathcal{P}^\epsilon_{\lambda(t,y)}[\hat{\Gamma}]$ including an initial density is defined  as
$
\mathcal{P}^\epsilon_{\lambda(t,y)}[\hat{\Gamma}] = \rho_0 (\Gamma(0)) \mathcal{P}^\epsilon_{\lambda(t,y)}[\hat{\Gamma}| \Gamma(0)]
$,
where $\rho_0(\Gamma)$ is the initial probability density at time $t=0$ for a transition . This path probability is assumed to be normalized by the path integral as $\int [\mathcal{D} \hat{\Gamma}] \mathcal{P}^\epsilon_{\lambda(t,y)}[\hat{\Gamma}] =1$. The probability density at time $t$ is defined as $\rho_t(\Gamma) = \int [\mathcal{D} \hat{\Gamma}] \delta(\Gamma(t)-\Gamma) \mathcal{P}^\epsilon_{\lambda(t,y)}[\hat{\Gamma}]$.
 For a feedback system, the ensemble averages of arbitrary path function $A[\hat{\Gamma}]$ and arbitrary phase function $B(\Gamma)$ are defined as
\begin{equation}
\left< A\right>_\epsilon= \int \prod_i dy_i \int [\mathcal{D} \hat{\Gamma}] A[\hat{\Gamma}] \mathcal{P}^\epsilon_{\lambda(t,y)}[\hat{\Gamma}]p_i(y_i|\Gamma_{M_i}),
\label{ensemble average 1}
\end{equation}
and
\begin{equation}
\left< B(t)\right>_\epsilon = \int \prod_i dy_i \int [\mathcal{D} \hat{\Gamma}]B(\Gamma(t))\mathcal{P}^\epsilon_{\lambda(t,y)}[\hat{\Gamma}]p_i(y_i|\Gamma_{M_i}).
\label{ensemble average 2}
\end{equation}
These ensemble averages are the averages for all paths and all measurement outcomes. To discuss the FDT violation, we define the response function $R(t;s)$ of the system for $t>s$, using the $\epsilon$ dependence of the ensemble average, as
\begin{equation}
\left< \dot{x}(t) \right>_\epsilon = \left< \dot{x}(t) \right>_0 + \epsilon \int^t_{0} ds R(t;s) f_p(s) +\mathrm{O}(\epsilon^2),
\label{response function}
\end{equation}
where $\left< \dots \right>_0$ is the ensemble average when the perturbation force $\epsilon f_p$ is $0$.
Due to the causality, $R(t,s)=0$ is satisfied for $t<s$. Moreover, the same time response is defined as $ R(t,t) = 1/2\left[R(t;t-0)+R(t;t+0)\right] =1/2R(t;t-0) $.

To consider a steady state, we generalize Eq.(\ref{SagawaUeda}) for several measurements and feedbacks. We define the $i$-th mutual information $I_i$ between the system's state $\Gamma_{M_{i}}$ and the measurement outcome $y_i$ as 
$I_i \equiv \ln p_i(y_i|\Gamma_{M_i})/p_i(y_i) $,
where $p_i(y_i)$ is the probability of obtaining the outcome $y_i$ in the $i$-th measurement. The probability $p_i(y_i)$ is calculated as
\begin{equation}
p_i(y_i)=\int \prod_{j \neq i} dy_j \int [\mathcal{D} \hat{\Gamma}] \mathcal{P}^\epsilon_{\lambda(t,y)}[\hat{\Gamma}] \Pi_{k}p_k(y_k|\Gamma_{M_k}).
\label{mutual information}
\end{equation}
Here, the normalization $\int dy_i p_i(y_i) = 1$ is satisfied.

\section{Main result}
For the Langevin system including feedback effects, we prove the inequality
\begin{eqnarray}
\beta \int^{\tau}_0 dt \gamma \left[ \left< \dot{x}(t)^2 \right>_0 - \frac{2}{\beta} R(t;t) \right] \geq \left< \Delta \phi \right>_0 - \sum_i \left<I_i \right>_0, \nonumber \\
\label{Main result 1}
\end{eqnarray}
where $\left< \Delta \phi \right>_0 =\left< \ln \rho_0(\Gamma(0)) -\ln \rho_{\tau}(\Gamma(\tau)) \right>_0 $ is the entropy change of the system.
This inequality shows that the time integral of the FDT violation is bounded by the sum of the mutual information. On the other hand, when the measurement outcome is obtained without error, the mutual information $\left< I_i \right>_0$ goes to infinity. In this limit, the bounds to the violation of FDT are vanished. When the measurement outcome is obtained with error, the mutual information has finite values. This result is interpreted as effects of error on the FDT violation. The bounds are crucial especially in the condition that the correlation term $\left< \dot{x}^2(t) \right>$ is smaller than the response term $2R(t;t)/\beta $. Therefore, the bounds can be important for a problem where the dynamical feedback makes the effective temperature of the system lower than the temperature of the heat reservoir because the effective temperature is defined as the ratio of the correlation term to the response term in a steady state,
\begin{equation}
T_{\rm eff} = \frac{\left< \dot{x}^2(t) \right>_0}{2 k_{\rm B} R(t;t)}.
\label{effective temperature}
\end{equation}
 The relation between the generalized FDT and the effective temperature is discussed in Ref.~\cite{Baiesi2}. When the system is considered to be in a steady state approximately by a time-independent feedback protocol as a result of coarse graining over time, we can prove the following relation for the effective temperature $T_{\rm eff}$
\begin{equation}
\frac{T_{\rm eff} -T}{T} \geq - \frac{\sum_i \left< I_i \right>_0}{\tau}t_r.
\label{Main result 2}
\end{equation}
where $t_r = m/\gamma$ is the relaxation time and $\sum_i \left< I_i \right>_0$ is the sum of the mutual information obtained within the time duration $\tau$. Then $\sum_i \left< I_i \right>_0/\tau$ is considered to be the mutual information rate obtained by the measurement. The left-hand side of Eq.(\ref{Main result 2}) is similar to the Carnot efficiency and the right-hand side is considered the information obtained in the relaxation time. It is worth indicating that $t_r$ is the characteristic time of the relaxation to an equilibrium state in terms of the velocity without external forces ($F_{\lambda} + \epsilon f_p=0$). In other words, the system practically forgets the information of the velocity after time $m/\gamma$. In order to cool the system down to a lower temperature, we should obtain the information of the velocity of a particle before the system loses the information of the velocity and applies feedback control. If the system is considered to be an overdamped Langevin system $(m/\gamma \to 0)$, the right-hand side of Eq.(\ref{Main result 2}) becomes $0$. Then this inequality indicates an inability to cool the overdamped Langevin system by the feedback force. In addition, if the relaxation time is smaller than the measurement interval, cooling the system is difficult and the lower bounds to the effective temperature are decided by this inequality. We prove these inequalities in the next section.

\section{Proof}
For the discussion of the detailed FT, a reversal process is introduced. Then we define a time-reversal map as $(x,\dot{x})^* = (x,-\dot{x})$. When $\hat{\Gamma}
$ is considered to be the trajectory of a forward process, the trajectory of the reverse process is described as $\hat{\Gamma^\dagger} = \{\Gamma^{*}(\tau-t)| 0 \leq t \leq \tau \}$. In the reversal process, a control parameter is introduced as $\lambda(\tau-t,y)$ using a protocol of the forward process $\lambda(t,y)$. We assume that the initial probability density of the trajectory of the reversed process is equal to the final probability density of the trajectory of the forward process ($\rho_0(\Gamma^*(\tau)) = \rho_{\tau}(\Gamma(\tau))$). According to Eq.(\ref{path integral expression}), the local detailed balance for the Langevin system is derived as
\begin{equation}
\frac{\mathcal{P}^\epsilon_{\lambda(t,y)}[\hat{\Gamma}]}{\mathcal{P}^\epsilon_{\lambda(\tau -t,y)}[\hat{\Gamma}^{\dagger}]} = \exp \left[ \int_0^{\tau} dt \omega(t) - \Delta \phi \right],
\label{local detailed balance}
\end{equation}
where $\omega(t)$ is the entropy production rate defined as
\begin{equation}
\omega(t) = \beta \dot{x}(t) \left[ F_{\lambda(t,y)}(x(t)) + \epsilon f_p(t)- m\ddot{x} (t) \right].
\label{entropy production rate}
\end{equation}
The entropy production rate $\omega(t) = \beta\dot{x}(t) \left[ \gamma \dot{x}(t) - \xi(t) \right]$ is consistent with the definition of the energy dissipation rate in Ref.~\cite{Sekimoto}.
The generalized Jarzynski equality~\cite{Sagawa2} for the system can be derived using the definition of the ensemble average including the feedback, Eq.(\ref{ensemble average 1}), as
\begin{eqnarray}
&&\left< e^{-\int_0^{\tau} dt \omega(t) + \Delta \phi - \sum_i I_i} \right>_\epsilon \nonumber \\
&=& \int \Pi_i dy_i p_i(y_i)\int [\mathcal{D} \hat{\Gamma}] \mathcal{P}^\epsilon_{\lambda(\tau-t,y)}[\hat{\Gamma}^{\dagger}] \nonumber \\
&=&1.
\label{generalized Jarzynski equality}
\end{eqnarray}
Due to the concavity of the exponential function, Jensen's inequality for Eq.(\ref{generalized Jarzynski equality}) is obtained.
 Then Jensen's inequality for $\epsilon = 0$ is equal to the generalization of the second law for a feedback Langevin system as
\begin{eqnarray}
\beta \int^\tau_0 dt \left<\dot{x}(t)\left[ F_{\lambda(t,y)}(x(t)) - m \ddot{x}(t) \right]  \right>_0 \nonumber\\
- \left< \Delta \phi \right>_0 \geq - \sum_i \left<I_i \right>_0,
\label{generalized second law}
\end{eqnarray}
because the left-hand side of Eq.(\ref{generalized second law}) is the entropy production from time $t=0$ to time $t=\tau$ and the entropy production is bounded by the sum of mutual information from time $t=0$ to time $t=\tau$.

To discuss the violation of FDT, we start with the identity
\begin{eqnarray}
&&\left. \frac{\partial}{\partial \epsilon}\left< \dot{x}(t) e^{-\epsilon \beta \int^\tau_0 dt' \dot{x}(t')f_p(t')} \right>_\epsilon \right|_{\epsilon=0}\nonumber \\
 &=&  \left. \frac{\partial \left< \dot{x}(t) \right>_\epsilon}{\partial \epsilon} \right|_{\epsilon=0}-\beta \int^\tau_0 dt' f_p(t') \left< \dot{x}(t) \dot{x}(t') \right>_0.
\label{identity 1}
\end{eqnarray}
The definition of the response function, Eq.(\ref{response function}), give us the relation
\begin{equation}
\left. \frac{\partial \left< \dot{x}(t) \right>_\epsilon}{\partial \epsilon} \right|_{\epsilon=0} = \int^t_{0} dt' R(t;t') f_p(t').
\label{response function 2}
\end{equation}
Moreover, we can calculate the identity, Eq.(\ref{identity 1}), exactly using the path probability, Eq.(\ref{path integral expression}), as
\begin{eqnarray}
&&\left. \frac{\partial}{\partial \epsilon}\left< \dot{x}(t) e^{-\epsilon \beta \int^\tau_0 dt' \dot{x}(t')f_p(t')} \right>_\epsilon \right|_{\epsilon=0}  \nonumber \\
&=& \frac{\beta}{2\gamma} \int^\tau_0 dt' f_p(t') \left< \dot{x}(t)\left[ -\gamma\dot{x}(t') -F_{\lambda(t',y)}(x(t')) \right. \right. \nonumber \\
&&\left. \left. +m\ddot{x}(t') \right] \right>_0. \label{identity 2}
\end{eqnarray}
A small impulse force $f_p(t')=\delta(t'-t+s)$ is substituted for Eqs.(\ref{identity 1})-(\ref{identity 2}) for $s \neq t$, then the generalized FDT for the feedback system can be derived as
\begin{eqnarray}
&&\gamma \left[ \left< \dot{x}(t) \dot{x}(t-s) \right>_0 - \frac{2}{\beta} R(t;t-s) \right] \nonumber \\
&=& \left< \dot{x}(t) \left[ F_{\lambda(t-s,y)}(x(t-s)) -m \ddot{x}(t-s) \right] \right>_0.
\label{Harada-Sasa}
\end{eqnarray}
using the causality $R(t;t+s)=0$ for $s>0$.
Owing to the definition of the Stratonovich-integral and the same time response $R(t;t)$, the relation between the same time response and correlation can be obtained as
\begin{eqnarray}
&&\gamma \left[ \left< \dot{x}^2(t) \right>_0 - \frac{2}{\beta} R(t;t) \right] \nonumber \\
 &=&\left< \dot{x}(t) \left[ F_{\lambda(t,y)}(x(t)) -m \ddot{x}(t) \right] \right>_0.
\label{Harada-Sasa 2}
\end{eqnarray}

This equality is the Harada-Sasa equality for a Langevin system with feedback. The left-hand side of Eq.(\ref{Harada-Sasa 2}) is the degree of violation of the FDT and the right-hand side of Eq.(\ref{Harada-Sasa 2}) represents the energy dissipation rate.  In an equilibrium state, the FDT violation is vanished because the feedback force $F_{\lambda(t,y)}(x(t))$ is considered to be a time-independent potential force, $-\partial U(x)/\partial x$. Therefore both correlations, $\left< \dot{x}(t) F_{\lambda(t,y)}(x(t)) \right>_0$ and  $\left< \dot{x}(t) \ddot{x}(t) \right>_0$ are $0$.

Therefore we obtain the first main result, Eq.(\ref{Main result 1}), from Eqs.(\ref{generalized second law}) and (\ref{Harada-Sasa 2}).
This result is valid for the Langevin dynamics driven by the feedback force. 

To discuss the effective temperature, we assume that the system is considered to be in a nonequilibrium steady state approximately as a result of coarse graining over time. A steady state can be introduced when the feedback protocol is independent of time. In our protocol, the $i$-th measurement outcome dependence of the feedback force is independent of $i$.
In a steady state, the correlation term $\left< \dot{x}^2(t) \right>$ and the response term $R(t;t)$ do not depend on time $t$. The effective temperature $T_{eff}$ is defined by the ratio of the correlation term to the response term in a steady state as Eq.(\ref{effective temperature}).
In an equilibrium state, the effective temperature is equal to the temperature of the heat reservoir because the degree of the FDT violation is 0, $\left< \dot{x}^2(t) \right>_0 - 2R(t;t)/\beta=0$,
while in a nonequilibrium steady state, the response function $R(t;t)$ is calculated using the Furutsu-Novikov-Donsker formula as in Refs.~\cite{Deutsch} and~\cite{Ohta} when the noise term $\xi(t)$ is a zero-mean white Gaussian noise. The correlation $\left< \dot{x}(t) \xi(t) \right>_0$ becomes $2\gamma R(t;t)/\beta$. Moreover, we can calculate $\left< \dot{x}(t) \xi(t) \right>_0$ by the definition of the Stratonovich integral.
When $\epsilon = 0$, the correlation $\left< \dot{x}(t) \xi(t) \right>_0$ is calculated as $\gamma/\left(m\beta\right)$.
The same time response $R(t;t)$ in a steady state is obtained exactly as $ R(t;t)=1/\left(2m\right)$.
This fact shows that the effective temperature fulfills 
\begin{equation}
\left< \frac{1}{2}m \dot{x}^2 \right>_0 = \frac{1}{2}k_{\rm B} T_{\rm eff}.
\label{effective temperature 2}
\end{equation}
If the probability of a particle's velocity is a zero-mean Gaussian distribution, Eq.(\ref{effective temperature 2}) means that the distribution of a steady state is considered to be the Maxwell-Boltzmann distribution with the temperature $T_{\rm eff}$. Let the value $R(t;t)=1/\left(2m\right)$, a steady-state condition $\left< \Delta \phi \right>_0 = 0$, and Eq.(\ref{effective temperature 2}) substitute for the first main result, Eq.(\ref{Main result 1}), then we can obtain the second main result, Eq. (\ref{Main result 2}).

\section{Models for cold damping}
First, we consider the cold damping process~\cite{Jourdan} or entropy pumping~\cite{Kim}, generally given by the following Langevin equation:
\begin{equation}
m\ddot{x}(t) +\gamma{x}(t) = -\gamma' \dot{x}(t) + \xi(t).
\end{equation}
In this model, $\gamma'$ is positive. This cold damping process was proposed in an experiment of cooling a Brownian particle by applying a velocity-dependent feedback $-\gamma' \dot{x}(t)$. In a realistic experimental setup, this feedback can be realized by using optical tweezers~\cite{Kim,Li}. In a steady state, the effective temperature of this system $T\gamma/\left(\gamma + \gamma'\right)$ was found to be lower than the temperature of the heat reservoir $T$. Thus this model is considered as the noise cancellation. The feedback of this model includes the velocity of the Brownian particle $\dot{x}(t)$ without a measurement error.

We substitute $F_{\lambda}=- \gamma' \dot{x}(t)$ into Eq.(\ref{Harada-Sasa 2}),
then the FDT violation of the system is calculated as
$- \gamma' \left< \dot{x}^2(t) \right>_0 - d/dt\left<\left(m/2\right) \dot{x}^2(t) \right>_0
$.
In a steady state, the condition $d/dt\left<\left(m/2\right) \dot{x}^2(t) \right>_0=0$ is derived because the term $\left<\left(m/2\right) \dot{x}^2(t) \right>_0$ does not depend on time $t$. Then the FDT violation $- \gamma' \left< \dot{x}^2(t) \right>_0$ is always negative in a steady state. The effective temperature of the system is calculated by the definition Eq.(\ref{effective temperature}) as $T_{\rm eff} = T\gamma/\left(\gamma + \gamma'\right)$. In the limit of $\gamma' \to \infty$, the effective temperature $T_{\rm eff}$ reaches 0 K. This model does not give the cooling bounds to the effective temperature by the mutual information because the feedback protocol is free of measurement errors, thus the mutual information goes to infinity. In terms of the measurement error, this model cannot describe the actual setup because the feedback protocol has measurement errors in the actual experimental setup. If the feedback protocol of the cold damping has measurement errors, the bounds to the FDT violation given by Eq.(\ref{Main result 1}) are dominant and therefore the effective temperature cannot reach 0 K. To discuss the effects of errors on the FDT violation, we consider the following two models including measurement errors. We show the validity of the bounds to the FDT violation given by Eq.(\ref{Main result 1}).

\subsection{Case 1}
A model for cold damping with continuous output feedback can be described by the Langevin equation
\begin{equation}
m\ddot{x}(t) + \gamma \dot{x}(t) = F_{\lambda(t,y)}(x(t)) + \xi(t).
\label{Feedback langevin eq 2}
\end{equation}
We consider the following feedback protocol for one cycle. First, a measurement about the velocity  $\dot{x}(0) = \dot{x}_0$ is performed at time $t=0$. Second, a measurement outcome $y$ about the velocity $\dot{x}_0$ is obtained. In order to introduce the measurement error, we consider that the conditional probability is Gaussian with variance $\sigma^2_{\rm err}$ as
\begin{equation}
p(y|\dot{x}_0) = \frac{1}{\sqrt{2\pi \sigma_{\rm err}^2}} \exp \left[ -\frac{(\dot{x}_0 - y)^2}{2\sigma_{\rm err}^2} \right].
\label{Condition probability}
\end{equation}
Third, a constant force $F_{\lambda(t,y)}(x(t)) = -\gamma' y$ is applied to the system from time $t=0$ to $t=\tau$.
This feedback sequence defines one cycle. In repeating this cycle, we assume that the system has the same Gaussian distribution about the velocity at time $t=0$ and $t=\tau$, instead of the assumption of a steady state, described as $p(\dot{x}_0) =p(\dot{x}(\tau)) = 1/\sqrt{2\pi \sigma^2} \exp \left[ - 
\dot{x}^2_0/\left(2 \sigma^2\right) \right].$ 
Due to the noise cancellation, the variance of the steady state density becomes smaller than that of the original Maxwell-Boltzmann distribution with temperature $T$ as $1/\left( m\beta \right) \geq \sigma^2.$

In this model, we can show the validity of Eq.(\ref{Main result 1}) for one cycle. 
Let the left hand side of Eq.(\ref{Main result 1}) be defined as $\Omega_{\tau} = \beta \int^{\tau}_0 dt \gamma \left[ \left< \dot{x}(t)^2 \right>_0 - 2R(t;t)/\beta \right]$.
The FDT violation, $\Omega_{\tau}$, can be calculated using Eq.(\ref{Harada-Sasa 2}) as
\begin{eqnarray}
\Omega_{\tau} &=& \beta \int^\tau_0 dt  \left< \dot{x}(t)F_{\lambda(t,y)}(x(t)) \right>_0 \nonumber\\
&&- \left< \beta \frac{m}{2} \left[ \dot{x}^2(0)-\dot{x}^2(\tau) \right] \right>_0,
\label{Calculation 1}
\end{eqnarray}
In this condition, the relations $\left< \Delta \phi \right>_0 =0$ and $\left< m/2\left[ \dot{x}^2(0)-\dot{x}^2(\tau) \right] \right>_0 =0$ are calculated because the probability distribution is the same at $t=0$ and $t=\tau$. Then we compare the value of the FDT violation $\Omega_{\tau}$ and the mutual information $\left<I \right>$ to discuss the validity of Eq.(\ref{Main result 1}). We can exactly calculate the violation of the FDT as
\begin{equation}
\Omega_{\tau} =- \beta \int^\tau_0 dt \int^\infty_{-\infty} dy \int^\infty_{-\infty} d\dot{x}_0 p(\dot{x}_0) p(y|\dot{x}_0) \gamma' y \bar{\dot{x}}(t),
\label{Calculation 2}
\end{equation}
where $\bar{\dot{x}}(t)$ is the average of the velocity in terms of the thermal noise $\xi(t)$. $\bar{\dot{x}}(t)$ obeys the equation of motion 
$
m\left(d/dt\right)\bar{\dot{x}}(t) = -\gamma \bar{\dot{x}}(t) - \gamma' y
$; then the solution of the equation of motion is calculated as
\begin{equation}
\bar{\dot{x}}(t) = - \frac{\gamma' y}{\gamma}+ \left( \dot{x}_0 + \frac{\gamma' y}{\gamma}  \right) e^{-\frac{\gamma}{m}t}.
\label{Calculation 4}
\end{equation}
Then we substitute Eqs.(\ref{Calculation 4}) and (\ref{Condition probability}) into Eq.(\ref{Calculation 2}) to obtain the value of the FDT violation as
\begin{eqnarray}
\Omega_{\tau} &=& \beta \frac{\gamma'^2}{\gamma}(\sigma^2+\sigma_{\rm err}^2)\tau \nonumber \\
&&- \beta \gamma' \frac{m}{\gamma} \left[ \sigma^2 + \frac{\gamma'}{\gamma} \left( \sigma^2+\sigma^2_{\rm err} \right) \right] \left( 1 -e^{-\frac{\gamma}{m} \tau} \right).
\label{Calculation 5}
\end{eqnarray}
When $\left(d \Omega_{\tau}/d\tau \right) \left. \right|_{\tau = \tau_{\rm min}} =0 $, the FDT violation has a minimum value in terms of $\tau$. The value of $\tau_{\rm min}$ is calculated as $\tau_{\rm min} = m/\gamma \ln \left[1 + \left(\gamma/\gamma'\right)\left[ \sigma^2/\left(\sigma^2 + \sigma^2_{\rm err}\right)\right] \right]$.
Therefore, the minimum value of the FDT violation $\Omega_{\tau_{\rm min}}$ is obtained as
\begin{eqnarray}
\Omega_{\tau_{\rm min}} &=& \frac{\beta m \gamma'^2(\sigma^2 + \sigma_{err}^2) }{\gamma^2} \ln \left(1 + \frac{\gamma}{\gamma'} \frac{\sigma^2}{\sigma^2 + \sigma_{err}^2} \right) \nonumber\\
&&- \frac{\beta m \gamma' \sigma^2}{\gamma} \nonumber\\
&\simeq& - \frac{m \beta \sigma^2}{2} \frac{1}{1+ \sigma_{r}^2}.
\label{Calculation 7}
\end{eqnarray}
where $\sigma_{r} = \sigma_{\rm err}/\sigma$.
In this calculation, the logarithmic term is expanded in terms of $\sigma^2/\left(\sigma^2 + \sigma_{\rm err}^2 \right)$ ($\leq 1$). 

On the other hand, the mutual information $\left< I \right>_0$ can be calculated. The probability of obtaining the measurement outcome $p(y)$ is calculated as
\begin{eqnarray}
p(y) &=& \int^{\infty}_{-\infty} d\dot{x}_0 p(y|\dot{x}_0)p(\dot{x}_0) \nonumber\\
 &=& \frac{1}{\sqrt{2 \pi (\sigma^2 + \sigma_{\rm err}^2)}} \exp \left[ - \frac{y^2}{2(\sigma^2+\sigma_{\rm err}^2)} \right].
\end{eqnarray}
Then, the mutual information $\left< I \right>_0$ is obtained as
\begin{eqnarray}
\left< I \right>_0 &=& \int^\infty_{-\infty} dy \int^\infty_{-\infty} d\dot{x}_0 p(\dot{x}_0) p(y|\dot{x}_0) \ln \frac{p(y|\dot{x}_0)}{p(y)} \nonumber \\
&=& \frac{1}{2} \ln \left( 1+ \frac{1}{\sigma_{r}^2} \right).
\label{Calculation 9}
\end{eqnarray}
The results of Eqs.(\ref{Calculation 7}) and (\ref{Calculation 9}) and the condition of the variance $1/\left(m\beta\right) \geq \sigma^2$ give us the inequality
\begin{equation}
\Omega_{\tau_{\rm min}} \geq - \frac{1}{2} \frac{1}{1+ \sigma_{r}^2} \geq -\left< I \right>_0 .
\label{Calculation 11}
\end{equation}
\begin{figure}
\includegraphics[scale=1.3]{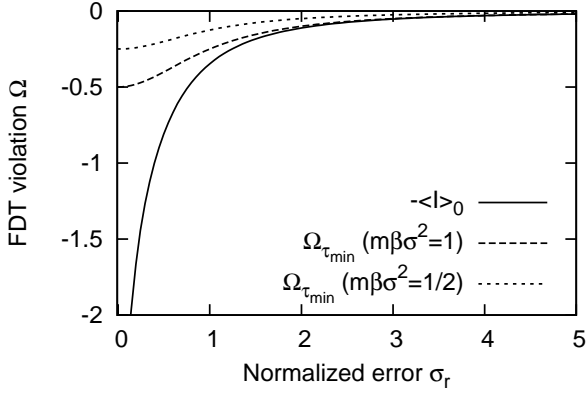}%
\caption{\label{fig.1} Minimum values of FDT violation (dashed lines) and mutual information (solid line) in case 1. Mutual information is less than the FDT violation. Thus the main result, Eq.(\ref{Main result 1}), is valid in case 1.}
\end{figure}

Thus the bounds to the FDT violation given by Eq.(\ref{Main result 1}) are valid in this model. According to Fig.~\ref{fig.1}, the bounds to the FDT violation are effective when the measurement error cannot be negligible ($\sigma_{r}^2 \gg 1$).
This result does not depend on any of the parameters $\tau$, $m$, $\gamma$, $\gamma'$, $\sigma$, or $\sigma_{\rm err}$. In other words, the validity of Eq.(\ref{Main result 1}) does not depend on the feedback parameter in this model.
\subsection{Case 2}
Here we consider the case where the output of feedback control system takes only discrete values. We assume that the system has only binary states for the measurement outcome. In such a case, without loss of generality, the measurement outcome $y$ can be simply represented by $y=0$ for negative values of $\dot{x}_0$ observations, or $y=1$ otherwise. The measurement error rate $q$ $(0 \leq q \leq 1/2)$ is introduced by the conditional probability as
\begin{eqnarray}
p(0|\dot{x}_0)=\left\{ \begin{array}{ll}
q & (\dot{x}_0 \geq 0) \\
1-q & (\dot{x}_0 < 0) \\
\end{array} \right.,
\label{case2-1}
\end{eqnarray}
\begin{eqnarray}
p(1|\dot{x}_0 )=\left\{ \begin{array}{ll}
1-q & (\dot{x}_0  \geq 0) \\
q & (\dot{x}_0  < 0) \\
\end{array} \right..
\label{case2-2}
\end{eqnarray}
Here, a constant force $F_{\lambda(t,0)}(x(t)) = \gamma'$ or $F_{\lambda(t,1)}(x(t)) = -\gamma'$ $(\gamma'>0)$
is applied to the system from time $t=0$ to time $t=\tau$, depending on the value of $y$.
This feedback sequence is considered as one cycle. In repeating this cycle, we also assume that the system has the same Gaussian distribution about the velocity at time $t=0$ and $t=\tau$ described as $p(\dot{x}_0) =p(\dot{x}(\tau)) = 1/\sqrt{2\pi \sigma^2} \exp \left[ - 
\dot{x}^2_0/\left(2 \sigma^2\right) \right].$ 
Moreover we also assume the condition $1/\left(m\beta\right) \geq \sigma^2.$ In this case, the violation of the FDT $\Omega_\tau$ can also be calculated as 
\begin{eqnarray}
\Omega_\tau = \beta \sum_y \int^\tau_0 dt \int^\infty_{-\infty} d\dot{x}_0 p(\dot{x}_0) p(y|\dot{x}_0) F_{\lambda(t,y)}(x(t)) \bar{\dot{x}}(t), \nonumber \\
\label{case2-3}
\end{eqnarray}
where $\bar{\dot{x}}(t)$ is calculated as
\begin{equation}
\bar{\dot{x}}(t) =\left. \pm \frac{\gamma'}{\gamma}- \left( - \dot{x}_0 \pm \frac{\gamma'}{\gamma}  \right) e^{-\frac{\gamma}{m}t} \right. .
\label{case2-4}
\end{equation}
The plus and minus signs in Eq.(\ref{case2-4}) correspond to $y=0$ and $y=1$, respectively. By substituting Eqs.(\ref{case2-1}), (\ref{case2-2}), and (\ref{case2-4}) into Eq.(\ref{case2-3}), we obtain the value of the FDT violation as
\begin{eqnarray}
\Omega_{\tau} = \beta \frac{\gamma'^2}{\gamma} \tau - \beta \frac{m}{\gamma}\left[ \frac{\gamma'^2}{\gamma} + 2(1-2q) \frac{\gamma' \sigma}{\sqrt{2\pi}} \right] \left( 1 -e^{-\frac{\gamma}{m} \tau} \right). \nonumber\\
\label{case2-5}
\end{eqnarray}
When $\left(d \Omega_{\tau}/d\tau \right) \left. \right|_{\tau = \tau_{\rm min}} =0 $, the FDT violation has its minimum value in terms of $\tau$. In this case, the value of $\tau_{\rm min}$ is calculated as $\tau_{\rm min} = m/\gamma \ln \left[ 1 + (1-2q) \left[ 2 \sigma \gamma/\left(\sqrt{2\pi}\gamma'\right)\right] \right]$.
Therefore, the minimum value of the FDT violation $\Omega_{\tau_{\rm min}}$ is obtained as
\begin{eqnarray}
\Omega_{\tau_{\rm min}} &=& \frac{\beta m \gamma'}{\gamma} \left[ \frac{\gamma'}{\gamma} \ln \left[ 1 + (1-2q)\frac{2\gamma\sigma}{\sqrt{2\pi}\gamma'} \right] \right. \nonumber\\
&&\left. - (1-2q)\frac{2\sigma}{\sqrt{2\pi}} \right] \nonumber\\
&\simeq& - \frac{m \beta \sigma^2}{\pi} (1-2q)^2.
\label{case2-6}
\end{eqnarray}
In this calculation, the logarithmic term is expanded in terms of ($1-2q$) ($\leq 1$). 

\begin{figure}
\includegraphics[scale=1.3]{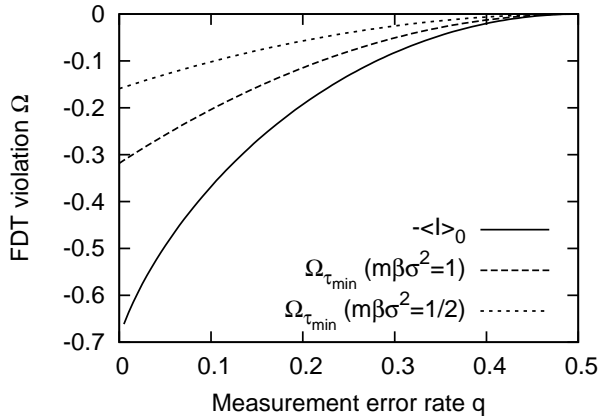}%
\caption{\label{fig.2} Minimum values of FDT violation (dashed lines) and mutual information (solid line) in case 2. Mutual information is less than the FDT violation. Thus the main result, Eq.(\ref{Main result 1}), is valid in case 2.}
\end{figure}

On the other hand, the probability of obtaining the measurement outcome $p(y)$ is calculated as $p(0) = 1/2$ and $p(1) = 1/2$. Then, the mutual information $\left< I \right>_0$ is obtained as
\begin{eqnarray}
\left< I \right>_0 &=& \sum_y \int^\infty_{-\infty} d\dot{x}_0 p(\dot{x}_0) p(y|\dot{x}_0) \ln \frac{p(y|\dot{x}_0)}{p(y)} \nonumber \\
&=& \ln 2 + q\ln q + (1-q) \ln (1-q)
\label{case2-7}
\end{eqnarray}
The results Eqs.(\ref{case2-6}) and (\ref{case2-7}) and the condition of the variance $1/\left(m\beta\right) \geq \sigma^2$ give us the inequality (see Fig.~\ref{fig.2})
\begin{equation}
\Omega_{\tau_{\rm min}} \geq - \frac{(1-2q)^2}{\pi} \geq -\left< I \right>_0 .
\end{equation}
Therefore we have demonstrated that Eq.(\ref{Main result 1}) is valid regardless of the different feedback protocols.
\section{Discussion}
In this paper, we have discussed the effects of error on the FDT violation and the effective temperature using Langevin dynamics under feedback with error. The bounds to the FDT violation and the effective temperature as a function of the mutual information are derived. Then we have presented two simple models to demonstrate analytical calculations for the validity of the generalized second law, Eq.(\ref{SagawaUeda}), for a Langevin system including the velocity-dependent feedback with error. Moreover, the result for the effective temperature is considered to be the relation between the information obtained by the measurement and the relaxation. We believe that this result is a valuable approach to nonequilibrium steady-state dynamics when the contents of the information play a significant role in feedback control systems.

As a possible experimental realization of the proposed results, cooling of a Brownian particle by application of a feedback force with laser tweezers might be a good candidate, since the velocity of a Brownian particle is measurable in the present technology~\cite{Li}. In vacuum, millikelvin  cooling of a Brownian particle was recently archived and the lowest temperature was limited by the noise~\cite{Li2}. For the generalized second law, the inequality, Eq.(\ref{SagawaUeda}), has been tested by our group in the feedback system of a Brownian particle~\cite{Toyabe}. Therefore experimental verification may be technically feasible. A more important extension of the present result will be the generalization to a quantum system, in which measurement error comes from quantum fluctuations, or generalization to many-particle systems. It is worth noting that the stochastic cooling in particle acceleration technology uses a periodic feedback control. It would be interesting to look for a theoretical relation with mutual information in many-particle systems as in Ref.\cite{vandermeer}.

\begin{acknowledgments}
\section{Acknowledgments}
The authors would like to thank Dr. T. Sagawa and Professor S. Sasa for their valuable comments.
\end{acknowledgments}

\end{document}